\documentclass[reprint, twocolumn, twoside,  nofootinbib, hidelinks, colorlinks, linkcolor=blue, citecolor=blue, urlcolor=black,
 amsmath,amssymb,
 aps, prx, 
floatfix,
]{revtex4-1}

\usepackage[switch,columnwise]{lineno}
\usepackage{fancyhdr}
\usepackage{helvet}
\usepackage{graphicx}
\usepackage{dcolumn}
\usepackage{bm}
\usepackage{hyperref}
\usepackage{mathrsfs}
\usepackage{amsbsy}
\usepackage{amsmath}
\usepackage[mathscr]{euscript}
\usepackage{scrextend}

\begin{document}

\preprint{APS/123-QED}

\title{Generalization of the Kelvin Equation for Arbitrarily Curved Surfaces}

\author{David V. Svintradze} 
 \email{d.svintradze@ug.edu.ge; dsvintra@yahoo.com}
\affiliation{%
 School of Health Sciences, The University of Georgia, 77a Kostava Str., Tbilisi 0171, Georgia
}%

\date{\today}

\begin{abstract}
Capillary condensation, which takes place in confined geometries, is the first-order vapor-to-liquid phase transition and is explained by the Kelvin equation, but the equations applicability for arbitrarily curved surface has been long debated and is a sever problem. Recently, we have proposed generic dynamic equations for moving surfaces. Application of the equations to static shapes and modelling the pressure at the interface nearly trivially solves the generalization problem for the Kelvin equation. The equations are universally true for any surfaces: atomic, molecular, micro or macro scale, real or virtual, Riemannian or \textit{pseudo}-Riemannian, active or passive. 
\end{abstract}

\maketitle


\textit{Introduction}. Capillary condensation is heterogeneous nucleation in nature \cite{2008635,PhysRevX.8.041046, YAROM2015743,F29868201763,doi:10.1063/1.113020,PhysRevLett.88.185505,Lee2015}. For instance when two hydrophilic surfaces are in contact, a nano-liquid meniscus is capillary condensed and it is playing an important role in adhesion, modulating kinetic friction and inducing cloud formation \cite{doi:10.1063/1.113020,PhysRevLett.88.185505,Lee2015}. To describe the process the macroscopic Kelvin equations was proposed \cite{Bocquet1998,FISHER1979}:
\begin{equation}
\frac{1}{R}=\frac{k_BT}{v_f\sigma}\ln\frac{p_v}{p_s} \label{kelvin}
\end{equation}
where $R$ is the mean radius so that $1/R=1/R_1+1/R_2...$, $k_B$ is the Boltzmann constant, $T$ is the temperature, $v_f$ is the molecular volume of the fluid, $p_v/p_s$ is the ratio of the external $p_v$ vapor pressure  to the $p_s$ saturation pressure  (which is same as the relative humidity) and $\sigma$ is the surface tension. The Kelvin equation (\ref{kelvin}) links the equilibrium curvature and the macroscopic parameters for fluid/vapor interface. 

While (\ref{kelvin}) was clarified to be accurate for spherical or near to spherical structures, its generalization for any arbitrarily curved surfaces remained sever problem. Recently it has been shown that the Kelvin equation holds down at nanometer scale when the curvature dependence of the surface tensions is taken into account \cite{PhysRevX.8.041046}. To be straightforward the curvature dependence of the surface tension was taken into account by generalizing the Kelvin equation for spherical or near to spherical structures by so called the Kelvin-Tolman equations. The last one can be easily obtained from the Kelvin equations if one proposes that the classically defined surface tension $\sigma$ is a function of the curvature, so that
\begin{equation}
\sigma=\frac{\sigma_0}{1+\frac{\delta}{R}} \label{tolman}
\end{equation}   
where $\delta$ is the Tolman length \cite{doi:10.1063/1.1747247}, $\sigma$ is the curvature dependent the surface tension and $\sigma_0$ is the surface tension of the planar interface. Taking (\ref{tolman}) modification into account (\ref{kelvin}) can be rewritten as
\begin{equation}
\frac{1}{R}=\frac{k_BT}{v_f\sigma_0}(1+\frac{\delta}{R})\ln\frac{p_v}{p_s} \label{kelvin-tolman}
\end{equation}
This generalization is the first order approximation of the curvature dependence which assumes some constant Tolman length \cite{doi:10.1063/1.1747247}. To be more accurate we can say that the Kelvin–Tolman equation is the same as the Kelvin equation with preposition that the last one holds for any curvature dependent surface tension. Though, the preposition that the Kelvin equation stands for any curvature dependant surface tension, for highly curved surfaces, has proven to be non trivial statement \cite{FISHER1979,doi:10.1021/la991404b,doi:10.1021/ja405408n}. Therefore, the generalization of the Kelvin equation for any arbitrarily curved surfaces is sever problem.

Recently, we have solved the dynamics problem of surfaces by deriving exact equations of motions for three and two-dimensional surfaces \cite{10.3389/fphy.2017.00037,10.3389/fphy.2018.00136,Svintradze2019}, see also \cite{svintradze2018shape, svintradze2017geometric, svintradze2016micelles, svintradze2016cell, svintradze2015moving, svintradze2014conformational, svintradze2013predictive, svintradze2011topology, svintradze2010hydrophobic, svintradze2009conformational}.\footnote{We cite conference abstracts here just to indicate, that speculations about existence of the generic equations of motions for moving surfaces we started at those abstracts and conferences. As a consequence, we reported initial version of such equations at the Biophysical Society's Annual Meeting 2015 \cite{svintradze2015moving}.} Application of the equations to static shapes and modelling the pressure at the interface nearly trivially solves the generalization problem. The Kelvin equation generalization problem is resolved without any prepositions that the Kelvin equation holds for every generic surface tensions. In fact, we show that the surface tension is not the key factor for any arbitrarily curved surfaces in chemical equilibrium contact. Instead, the Kelvin equation is the specific case of the equation we present here and holds only when the surface is homogeneous, has the time invariable surface tension and is in equilibrium with the environment.


\textit{Equations of Motions}. To make the paper self-sustained, we give brief introduction to the covariant equations of motions for moving surface and basic principles behind the derivation. 

Definitions of metric tensor, base vectors, the surface velocity in the ambient space and theorems needed for derivations of the equations of motions, also basics of the Riemannian geometry and its extension to the moving surfaces can be found in our recent papers \cite{10.3389/fphy.2017.00037,10.3389/fphy.2018.00136,Svintradze2019}. 

The surface base vectors are defined as partial derivative of the position vector $\bm{R}$ so that $\bm{S_a}=\partial_a\bm{R}$. Vectors are designated as bold letters throughout the paper and the summation convention follows to the Einstein convention, repeated upper and down indexes indicate the summation by the index. Latin letters in the indexes display tensors related to the surface and as far as here we deal with two dimensional surfaces the Latin indexes run though one and two ($a=1,2$). Greek indexes are related to the tensors defined on the space and are natural numbers up to three ($\alpha=1,2,3$). 

The surface metric tensor is a dot product of the base vectors of the tangent plane $g_{ab}=\bm{S_a\cdot S_b}$ and its velocity is defined as sum of the normal $C$ and tangent $V^a$ velocities so that 
\begin{equation}
\bm{V}=C\bm{N}+V^a\bm{S_a} \label{velocity}
\end{equation}
here $N$ is the surface normal 
Since we deal with dynamic surfaces all parameters: base vectors, velocities, metric tensor, the surface area $S$, topology and enclosed volume $\Omega$ are functions of parametric time $t$.  Note, that since the surface velocity is the ambient one-tensor, generally it also can be defined as the time derivative of the position vector and can be represented in ambient space base vectors $\bm{X_\alpha}$ as $\bm{V}=V^\alpha \bm{X_\alpha}$ where $V^\alpha=\partial X^\alpha/\partial t$ is the ambient $\alpha$ component of the velocity.

The definitions of base vectors, metric tensor and the surface velocity form the core principle for defining curvilinear invariant derivatives and extension to the invariant time derivative, so that for any arbitrarily defined tensor $T_a^b$ the following stands
\begin{equation}
\dot\nabla T_a^b=\frac{\partial T_a^b}{\partial t}-V^n\nabla_nT_a^b+\dot\Gamma^b_nT^n_a-\dot\Gamma^n_aT_n^b \label{time derivative}
\end{equation} 
where $\dot\Gamma_a^b=\nabla_aV^b-CB_a^b$ is the so called Christoffel symbol for the moving surface and $B_{ab}=\bm{N}\cdot\nabla_a\bm{S_b}$ is the curvature tensor \cite{10.3389/fphy.2017.00037,10.3389/fphy.2018.00136}. Along with invariant time derivative there is a theorem for taking time derivative for the space integral. For any scalar field $f$, defined on the $\Omega$ space surrounded by the $S$ moving surface, the following theorem stands:
\begin{equation}
\frac{d}{dt}\int_\Omega fd\Omega=\int_\Omega\frac{\partial f}{\partial t}d\Omega+\int_SfCdS \label{time deriv integral}
\end{equation}  
Consequently, in a case of the compact space with conserved volume the theorem (\ref{time deriv integral}) dictates $C=0$ condition.
 
Above definitions form fundamental principles of calculus for moving surfaces and provide basic tools for straightforward derivation of the surface dynamic equations. We have provided exact derivation few times before \cite{10.3389/fphy.2017.00037,10.3389/fphy.2018.00136}. To avoid self-repetition but give an introduction to the equations generality, we provide generic equations for two-dimensional surface dynamics here and give only basics to the derivation. We start from the generic Lagrangian of the surface motion
\begin{equation}
\mathcal{L}=\int_S\frac{\rho V^2}{2}dS-\int_\Omega ud\Omega \label{lagrangian}
\end{equation} 
where $u$ is the density of the potential field on $\Omega$ and $\rho$ is the surface mass density. For convenience note that the $u$ has the same dimension as a pressure and in fact, for infinitesimal volumes, it is the same as the negative internal surface pressure $p$ applied by the $\Omega$ space to the $S$ surface: $p_{int}=-u$ for internal pressure or $p_{ext}=u$ if it is the external pressure applied by the environment. If the interaction with the environment has to be taken into account then the potential field becomes sum of the internal and external fields and the total surface pressure becomes difference between external and internal pressures 
\begin{equation}
p=p_{ext}-p_{int} \label{ext-int}
\end{equation}
 
Mass balance dictates that at the absence of shape dynamics the surface mass must be conserved. Note that the boundary condition dictated by the conservation of mass does not demand that the surface must be initially massive nor it must have constant mass. The boundary condition $\frac{d}{dt}\int_S\rho dS=0$ lands the generalization of continuity equation, reading:
\begin{equation}
\dot\nabla\rho+\nabla_a(\rho V^a)=\rho CB_a^a \label{mass balance}
\end{equation}
where $B_a^a$ is the trace of the mixed curvature tensor and is the mean curvature (see detail derivation in papers \cite{10.3389/fphy.2017.00037,10.3389/fphy.2018.00136}. This generalization of continuity equation has been reported and successfully used in various applications before, see for instance \cite{PhysRevLett.105.137802} and references therein.

For the variation of the space integral we note: the potential energy part of the (\ref{lagrangian}) Lagrangian, follows the theorem for the space integration and can be calculated as
\begin{equation}
\delta\int_\Omega pd\Omega=\int_\Omega\frac{\partial p}{\partial t}d\Omega+\int_SpCdS \label{integral p}
\end{equation}
If the system is incompressible then only the first term from the right hand side of the equation (\ref{integral p}) survives. The second term of (\ref{integral p}) is the normal variation and by the Gauss theorem can be converted to the space integral as
\begin{equation}
\int_S pCdS=\int_S pV^\alpha N_\alpha dS=\int_\Omega \partial_\alpha (pV^\alpha) d\Omega \label{p normal}
\end{equation} 
(\ref{p normal}) vanishes when the system is incompressible. The first term of the (\ref{integral p}) has normal component, which can be modelled as $V^\alpha\partial_\alpha p$, and tangent terms modelled as $V_iN^\alpha\nabla^ip_\alpha$ \cite{10.3389/fphy.2017.00037,10.3389/fphy.2018.00136}. The variation of the kinetic part was a tricky and required extension of differential geometry to account moving surfaces. The normal component of the $\delta\int_S\rho V^2/2dS$ variation is $\int_S\rho C(\dot\nabla C+2V^i\nabla_iC+V^aV^bB_{ab})dS$ and tangent components come from the integral $\int_S\rho V_i(\dot\nabla V^i+V^j\nabla_jV^i-C\nabla_iC-CV^jB_j^i)dS$. According to the minimum action principle the normal and tangent components of the kinetic energy variation must be identical to the normal and tangent parts of the potential energy variation, therefore taking into account  (\ref{mass balance}) we end up with the equations of motions:
\begin{align}
&\dot{\nabla}\rho+\nabla_i(\rho V^i)=\rho CB_a^a \nonumber  \\
&\partial_\alpha(V^\alpha(\rho(\dot{\nabla} C+2V^i\nabla_iC+V^iV^jB_{ij})+p))=-V^\alpha\partial_\alpha p \nonumber \\
&\rho(\dot{\nabla}V^i+V^a\nabla_aV^i-C\nabla^iC-CV^jB_j^i)=-N^\alpha\nabla^ip_\alpha \label{eq motions}
\end{align}
The equations (\ref{eq motions}) are complete set for the surface dynamics, as far as have four unknowns $\rho,C, V_1,V_2$ and four differential equations. All information, about how the internal processes may effect on the surface dynamics, is stored in the surface pressure term, which can be subject of the modelling dependently on the nature of the problem. Because the Lagrangian (\ref{lagrangian}) is invariant and the variation is taken by tensor calculus, the equations are fully covariant.


\textit{Solution}. We now show that the equations of motions (\ref{eq motions}) nearly trivially provide the solutions for the Kelvin equation of arbitrarily curved surfaces. Indeed, lets assume incompressible liquid in contact with the vapor. For simplicity we provide equations for incompressible fluids, though generalization for compressible ones is not conceptually difficult. Incompressibility condition dictates conservation of the volume so that the interface velocity $C$, that is normal velocity of the surface, must be zero. This follows from the fact that the volume motion is associated to the surface normal motion according to (\ref{time deriv integral}). The condition $C=0$ dictates that divergence of the surface velocity must also vanish $\partial_\alpha V^\alpha=0$. These two conditions simplify the equations for the surface normal motion, so that:
\begin{equation}
\rho V^iV^jB_{ij}=- p \label{pressure}
\end{equation}
Before we proceed further note that the term $\rho V^iV^j$ is the density of the kinetic energy stress tensor, therefore the tensor
\begin{equation}
T^{ab}=V^aV^b \label{stress tensor}
\end{equation}
causes deformations of the surface in $a,b$ directions. With these notations the equation (\ref{pressure}) becomes
\begin{equation}
\rho T^{ab}B_{ab}=- p \label{stress-pressure}
\end{equation}
Since we deal with the fluid/vapor interface, according to (\ref{ext-int}) the surface pressure can be modelled as $p=p_v-p_f$, where $p_v,p_f$ stand for vapor pressure and fluid pressure at the interface respectively. 

Now, lets assume the incompressible liquid in contact with the vapor, satisfying ideal gas law, and the transition from planar surface to a curved one goes in chemically equilibrated process. Chemical equilibrium dictates that the change of the chemical potential of the vapor $\Delta\mu_v$ must be equal to the change of the chemical potential of the fluid $\Delta\mu_f$ while the interface curves
\begin{equation}
\Delta\mu_v=\Delta\mu_f \label{equi}
\end{equation}
According to Gibbs-Duhem equation $d\mu=vdp-sdT$ where $s$ is the entropy, $T$ is the temperature, $v$ is the volume and $p$ is the pressure. Therefore, the change of chemical potentials of the vapor and the fluid are
\begin{equation}
\Delta\mu_v=\int_{p_s}^{p_{v}}\frac{k_BT}{p}dp=v_f\int_{p_s}^{p_f}dp=\Delta\mu_f \label{chem equ}
\end{equation}  
here $p_s$ is the saturation pressure and $p_v, p_f$ are external vapor and internal fluid pressures at the interface, $v_f$ is the molecular volume of the liquid. Using (\ref{ext-int}) and (\ref{stress-pressure}) in the chemical potential for the fluid (\ref{chem equ}), we end up with
\begin{align}
v_f\int_{p_s}^{p_f}dp&=v_f(p_f-p_s)=v_f(p_f-p_v+p_v-p_s)\nonumber \\ 
&=v_f(\rho T^{ab}B_{ab}+p_v-p_s)=k_BT\ln\frac{p_v}{p_s} \nonumber
\end{align}
Since, for highly curved surfaces $\rho T^{ab}B_{ab}>>p_v-p_s$, the last equation lands generic equation 
\begin{equation}
T^{ab}B_{ab}=\frac{k_BT}{M}\ln H \label{GK}
\end{equation}
where $M=\rho v_f$ is the molar mass of the fluid/vapor interface (the surface) and $H=p_v/p_s$ is the relative humidity. The equation (\ref{GK}) is the generalization of the Kelvin equation for arbitrarily curved surfaces.

\textit{Relevance to the Kelvin Equation}. Now we can argue that (\ref{GK}) is indeed the generalization we were looking for. To prove that we just need to indicate that in some limits (\ref{GK}) simplifies to the Kelvin and the Kelvin-Tolman equations.  

Indeed, lets assume that the surface is homogeneous and can be described by time invariable surface tension $\sigma$, then as we have proved before
\begin{equation}
\rho(\dot{\nabla} C+2V^i\nabla_iC+V^iV^jB_{ij})=\sigma B_a^a \label{normal tension}
\end{equation}
For the proof see the papers \citep{10.3389/fphy.2017.00037, 10.3389/fphy.2018.00136}. Simplifying the equation (\ref{normal tension}) with the condition $C=0$ and taking into account (\ref{stress tensor}), we get  $\rho V^iV^jB_{ij}=\sigma B_a^a$ and therefore (\ref{GK}) transforms into
\begin{equation}
\sigma B_a^a=\frac{k_BT}{v_f}\ln\frac{p_v}{p_s} \label{mean curvature Kelvin}
\end{equation}
Taking into account that for spherical surface, with sign convention, $B_a^a=1/R$ (where $R$ is the mean radius so that $1/R=1/R_1+1/R_2$), then (\ref{mean curvature Kelvin}) becomes exactly the Kelvin equation (\ref{kelvin}). 

Note, that in the (\ref{mean curvature Kelvin}), even though the surface tension is time invariable, it can be a function of the mean curvature $\sigma=\sigma(B_a^a)$. Therefore, expansing $\sigma(B_a^a)$ by Taylor series, in the first approximation, one will get (\ref{tolman}) and substitution it to (\ref{mean curvature Kelvin}) leads to the Kelvin-Tolman equation (\ref{kelvin-tolman}). Thereby, we rigorously explain why the the Kelvin-Tolman equation (\ref{kelvin-tolman}) has been successfully used in the recent nano-scale experiments \cite{PhysRevX.8.041046}. 

\begin{acknowledgements}
I would like to thank Max Planck Institute for the Physics of Complex Systems for the hospitality and thank Dr. Julicher (MPIPKS), Dr. Frey (LMU), Dr. Grosberg (NYU) and Dr. Arovas (UCSD) for discussions about moving surfaces. The work was initiated at the Aspen Center for Physics in 2017, which is supported by National Science Foundation grant PHY-1607611. My presence at the  center was supported by Simons Foundation. 
\end{acknowledgements}

\bibliographystyle{apsrev4-1} 
\bibliography{Ref1}

\end{document}